\documentstyle[12pt]{article}

\begin{document}

\sloppy

\begin{flushright}{UT-713}\end{flushright}

\vskip 1.5 truecm

\centerline{\large{\bf Tadpole Method and Supersymmetric}}
\centerline{\large{\bf O(N) Sigma Model}}
\vskip .75 truecm
\centerline{\bf Tomohiro Matsuda}
\vskip .4 truecm
\centerline {\it Department of Physics, University of Tokyo}
\centerline {\it Bunkyo-ku, Tokyo 113,Japan}
\vskip 1. truecm

\makeatletter
\@addtoreset{equation}{section}
\def\theequation{\thesection.\arabic{equation}}
\makeatother

\vskip 1. truecm

\begin{abstract}
We examine the phase structures of the supersymmetric O(N)
sigma model in two and three dimensions by using the tadpole
method.
Using this simple method, the calculation is largely
simplified and the characteristics of this theory
become clear.
We also examine the problem of the fictitious negative energy state.
\end{abstract}

\section{Introduction}
\hspace*{\parindent}
Many years ago, Gross and Neveu\cite{gn}
have shown that dynamical symmetry break
down is possible in asymptotically free field theories.
They obtained an expansion in powers of $1/N$ that is non-perturbative
in $g^{2}$.
This leads to a massive fermion and to a $\overline{\psi}{\psi}$
bound state at threshold.

Polyakov\cite{sigma}
 has pointed out that the $O(N)$  sigma model is 
asymptotically free and that the fundamental particle acquires a mass
for $N>2$.

Witten \cite{witten1}
has constructed a supersymmetric version of the two-dimensional
O(N) sigma model.
This is a hybridization of the non-linear sigma model and Gross-Neveu
model with Majorana fermions. 

There comes a natural question:
What is the difference between non-supersymmetric models and
supersymmetric ones?
If there is any difference, how is it realized?
Many authors tried to answer this question\cite{davis,alv}, 
but some questionable arguments are still left. 
The problem of the negative energy state is one of them\cite{maha,kazama}. 
To maintain the positivity of the vacuum energy, inclusion of 
the chiral condensation effect was crucial in ref\cite{kazama}.
However in the three dimensional model there is a weak coupling phase
where the chiral condensation vanishes but the bosonic condensation
is still possible.

The purpose of this paper is to clarify these ambiguities and present
a systematic treatment of this model.
To show explicitly what is going on, we are not going to use the 
equation 
of motion for supersymmetric auxiliary fields at the first stage. 
If we eliminate these fields, it becomes difficult to find what 
relations we are dealing with.

\section{Review of the non-linear sigma model}
\hspace*{\parindent}
In this and the next section we are going to review well-known
results on the O(N) non-linear sigma model and the four-fermion model for
the convenience of checking the notations.
If readers feel boring, please skip to section 4.

The Lagrangian for the O(N) sigma model is defined by
\begin{equation}
  L=-\frac{1}{2}n_{j}\partial^{2}n_{j}
\end{equation}
with the  local non-linear constraint
\begin{equation}
  n_{j}n_{j}=\frac{N}{g^{2}}.
\end{equation}
The sum over the flavor index j runs from 1 to N.
This constraint can be implemented by introducing a Lagrange 
multiplier $\lambda$.

Let us consider the Euclidean functional integral in the form:
\begin{eqnarray}
  Z&=&\int{D}\vec{n}\delta\left((\vec{n})^{2}-\frac{N}{g^{2}}\right)
  exp\left(-\frac{1}{2}\int(\partial_{\mu}\vec{n})^{2}d^{D}x
  \right)\nonumber\\
  &=&\int{D}\lambda\int{D}\vec{n}exp\left(-\frac{1}{2}\int\left\{
  (\partial_{\mu}\vec{n})^{2}+\lambda\left((\vec{n})^{2}-
  \frac{N}{g^{2}}\right)\right\}d^{D}x\right)
\end{eqnarray}
The integral over $n$ is Gaussian and can be performed in a 
standard fashion. 
We have:
\begin{equation}
  Z=\int{D}\lambda{exp}\left(\frac{N}{2g^{2}}\int{\lambda}d^{D}x
  -\frac{N}{2}trln(-\partial^{2}+\lambda)\right)
\end{equation}
Let us first compute the variation of the action with respect to 
$\lambda$.
We get\cite{polyakov}:
\begin{eqnarray}
  \label{gap1}
  \frac{N}{2g^{2}}&=&\frac{N}{2}\frac{\delta}{\delta\lambda}trln
  (-\partial^{2}+\lambda)\nonumber\\
  &=&\frac{N}{2}G(x,x;\lambda)
\end{eqnarray}
Here we have introduced the Green function:
\begin{equation}
  G(x,y;\lambda)=<y|(-\partial^{2}+\lambda)^{-1}|x>
\end{equation}
The meaning of the above equation becomes transparent if we
notice that
\begin{eqnarray}
  <n_{i}(x)n_{j}(y)>&=&Z^{-1}\int{D}\lambda\int{D}\vec{n}
  exp\left(-\frac{1}{2}\int\left\{(\partial\vec{n})^{2}+
  \lambda\left(\vec{n}^{2}-\frac{N}{g^{2}}\right)\right\}d^{D}x
  \right)\nonumber\\
  && \ \ \times{n}_{i}(x)n_{j}(y)\nonumber\\
  &=&\delta_{ij}\frac{\int{D}\lambda{e}^{W}G(x,y;\lambda)}
  {\int{D}\lambda{e}^{W}}\\
  W&=&\frac{N}{2g^{2}}\int\lambda{d}^{D}x-\frac{N}{2}trln
  (\partial^{2}+\lambda)\nonumber.
\end{eqnarray}
If $\lambda$ integration is to be approximated by the saddle point
$\lambda_{0}$, we obtain
\begin{equation}
   <n_{i}(x)n_{j}(y)>=\delta_{ij}G(x,y;\lambda_{0}).
\end{equation}
These equations show that eq.(\ref{gap1}) is nothing but the 
condition $<\vec{n}^{2}>=\frac{N}{g^{2}}$.
This is the main idea of the tadpole method\cite{tad}.
Let us now solve eq.(\ref{gap1}).
Passing to the momentum representation,
\begin{eqnarray}
  \label{gap2}
  G(x,y;\lambda_{0})&=&\int\frac{d^{D}p}{(2\pi)^{D}}
  \frac{e^{ip(x-y)}}
  {p^{2}+\lambda_{0}}\nonumber\\
  \frac{N}{2g^{2}}&=&\frac{N}{2}G(x,x;\lambda_{0})\nonumber\\
  &=&\frac{N}{2}\int\frac{d^{D}p}{(2\pi)^{D}}
  \frac{1}{p^{2}+\lambda_{0}}.
\end{eqnarray}
For D=2 we obtain:
\begin{eqnarray}
  1&=&\frac{g^{2}}{4\pi}log\frac{\Lambda^{2}}{\lambda_{0}}
  \nonumber\\
  \lambda_{0}&=&\Lambda^{2}exp\left(-\frac{4\pi}{g^{2}}\right)
\end{eqnarray}
For D=3, the situation is slightly different.
We have a critical coupling $g^{2}_{cr}$ defined by
\begin{equation}
  1=g^{2}_{cr}\int\frac{d^{3}p}{(2\pi)^{3}}\frac{1}{p^{2}}.
\end{equation}
If $g^{2}>g^{2}_{cr}$ then the equation has a non-trivial solution
$\lambda_{0}\ne0$.
Using $g_{cr}$, we can rewrite (\ref{gap2}) as:
\begin{eqnarray}
  \label{gap3}
  1&=&g^{2}\int\frac{d^{3}p}{(2\pi)^{3}p^{2}}-g^{2}\int
  \frac{d^{3}p}{(2\pi)^{3}}\left(\frac{1}{p^{2}}-\frac{1}{p^{2}
      +\lambda_{0}}\right)\nonumber\\
    &=&\frac{g^{2}}{g^{2}_{cr}}-g^{2}\int\frac{d^{3}p}{(2\pi)^{3}}
    \frac{\lambda_{0}}{p^{2}(p^{2}+\lambda_{0})}
\end{eqnarray}
The integral in (\ref{gap3}) is convergent and proportional
to $\lambda^{\frac{3}{2}-1}=\sqrt{\lambda}$.
Therefore, we have:
\begin{equation}
  m_{n}^{2}\equiv\lambda_{0}=const.\Lambda^{2}
  \left(\frac{g^{2}-g^{2}_{cr}}
  {g_{cr}^{2}}\right)^{2}
\end{equation}
If we take $g^{2}<g_{cr}^{2}$ something goes wrong with (\ref{gap3}).
It does not have any solution, so the constraint 
$<\vec{n}^{2}>=\frac{N}{g^{2}}$
cannot be satisfied.

We should also consider the possibility of spontaneous breaking 
of O(N) symmetry.
In above discussions, we have implicitly assumed that the 
vacuum expectation value of $\vec{n}$ would vanish.
Let us consider what may happen if $\vec{n}$ itself gets non-zero
vacuum expectation value.
Because of O(N) symmetry, the vacuum expectation value of
$\vec{n}\equiv(n_{1},n_{2},...n_{N})$ may be written as
\begin{equation}
  <\vec{n}>=(0,0,...\sqrt{N}v/g).
\end{equation}
So that the constraint equation (\ref{gap1}) becomes
\begin{eqnarray}
  \label{gapv}
  <(\vec{n})^{2}>&=&<\vec{n}>^{2}+<1-loop>\nonumber\\
  &=&N\left(\frac{v^{2}}{g^{2}}+\int\frac{d^{3}p}{(2\pi)^{3}}
  \frac{1}{p^{2}+\lambda_{0}}\right)\nonumber\\
  &=&\frac{N}{g^{2}}.
\end{eqnarray}
Of course, in two dimensions we cannot expect $\vec{n}$
to get any expectation value.
For D=3, we have another critical coupling $g'_{cr}$:
\begin{equation}
  \label{v}
  \frac{1-v^{2}}{g_{cr}^{'2}}=\int\frac{d^{3}p}{(2\pi)^{3}}
  \frac{1}{p^{2}}
\end{equation}
If $g$ is smaller than $g_{cr}$, then $v$ grows. As a result, 
the constraint equation has a solution in the weak coupling
 region($g'_{cr}\leq{g}\leq{g}_{cr}$)
 in a sense that not eq.(\ref{gap2}) but eq.(\ref{gapv})
is satisfied by some $\lambda_{0}$. 
As far as we are dealing with the non-supersymmetric sigma model,
we have no primary reason to believe that the vacuum 
expectation value
of the field $v=<n_{j}>$ would not obtain a non-vanishing 
value in the strong coupling region in three dimensions.

\section{Review of the four-fermion model}
\hspace*{\parindent}
The four-fermion model is described by the Lagrangian
\begin{equation}
  \label{lag4}
  L=\frac{i}{2}\overline{\psi}_{j}\not{\! \partial}\psi_{j}
  +\frac{g^{2}}{8N}(\overline{\psi}_{j}\psi_{j})^{2}
\end{equation}
where the sum of the flavor index j runs from 1 to N and we
require that $g^{2}$ remains constant as N goes to infinity.
By introducing a scalar auxiliary field $\sigma$ we may rewrite 
(\ref{lag4}) as
\begin{equation}
  L=\frac{i}{2}\overline{\psi}_{j}\not{\! \partial}\psi_{j}
  +\frac{1}{2}\sigma\overline{\psi}_{j}\psi_{j}-\frac{N\sigma^{2}}
  {2g^{2}}.
\end{equation}
Let us consider the functional integral in the form:
\begin{equation}
  Z=\int{D}\psi_{j}D\sigma{exp}\left[\int{d}^{D}x\left\{
  \frac{1}{2}\overline{\psi}_{j}(i\not{\! \partial}+\sigma)\psi_{j}
  -\frac{N}{2g^{2}}\sigma^{2}\right\}\right]
\end{equation}
Integrating over the field $\psi_{j}$ we get an effective action
for the field $\sigma$:
\begin{equation}
  Z=\int{D}\sigma{exp}\left[-\frac{N}{2g^{2}}\int{d}^{D}x
  \sigma^{2}+\frac{N}{2}Trln(i\not{\! \partial}+\sigma)\right]
\end{equation}
We impose the stationary condition which gives the gap equation.
\begin{equation}
  \frac{N<\sigma>}{g^{2}}-\frac{N}{2}\int\frac{d^{D}p}{(2\pi)^{D}}
  tr\frac{1}{-\not{\! p}+<\sigma>}=0
\end{equation}
As is in the non-linear sigma model discussed in the previous section,
this represents the condition
\begin{equation}
  \frac{N}{g^{2}}<\sigma>=
    \frac{1}{2}<\overline{\psi}_{j}\psi_{j}>|_{m_{\psi}=<\sigma>}.
\end{equation}
For D=2 we obtain:
\begin{eqnarray}
  \frac{1}{g^{2}}&=&\int\frac{d^{2}p}{(2\pi)^{2}}
  \frac{1}{p^{2}+<\sigma>^{2}}\nonumber\\
  <\sigma>^{2}&=&\Lambda^{2}exp\left(-\frac{4\pi}{g^{2}}\right)
\end{eqnarray}
For D=3, we have a critical coupling constant.
The saddle point exists only within the branch
\begin{equation}
  0<\frac{1}{g^{2}}\leq\frac{1}{g_{cr}^{2}}
\end{equation}
where
\begin{equation}
  \frac{1}{g^{2}_{cr}}
  \equiv\int\frac{d^{3}p}{(2\pi)^{3}}\frac{1}{p^{2}}.
\end{equation}

\section{Phases in the Supersymmetric Non-Linear Sigma Model}
\hspace*{\parindent}
The supersymmetric non-linear sigma model is usually defined by
the Lagrangian
\begin{equation}
  L=\frac{1}{2}\int{d}^{2}\theta\Phi_{j}D^{2}\Phi_{j}
\end{equation}
with the non-linear constraint
\begin{equation}
  \label{const2}
  \Phi_{j}\Phi_{j}=\frac{N}{g^{2}}.
\end{equation}
where the sum of the flavor index j runs from 1 to N.
The superfields $\Phi_{j}$ may be expanded out in components
\begin{equation}
  \Phi_{j}=n_{j}+\overline{\theta}\psi_{j}+\frac{1}{2}
  \overline{\theta}\theta{F}_{j}
\end{equation}
and the super covariant derivative is 
\begin{equation}
  D=\frac{\partial}{\partial\theta}-i\overline{\theta}
  \not{\! \partial}.
\end{equation}
In order to express the constraint (\ref{const2}) as a $\delta$
function, we introduce a Lagrange multiplier superfield $\Sigma$.
\begin{equation}
  \Sigma=\sigma+\overline{\theta}\xi+\frac{1}{2}\overline{\theta}
  \theta\lambda
\end{equation}
We thus arrive at the manifestly supersymmetric action for the 
supersymmetric sigma model.
\begin{equation}
  \label{lag3}
  S=\int{d}^{D}xd^{2}\theta\left[\frac{1}{2}\Phi_{j}D^{2}\Phi_{j}
  +\frac{1}{2}\Sigma\left(\Phi_{j}\Phi_{j}-\frac{N}{g^{2}}
  \right)\right]
\end{equation}
In component form, the Lagrangian from (\ref{lag3}) is
\begin{eqnarray}
  L&=&-\frac{1}{2}n_{j}\partial^{2}n_{j}+\frac{i}{2}
  \overline{\psi}_{j}\not{\! \partial}\psi_{j}+\frac{1}{2}
  F_{j}^{2}
  -\sigma{n}_{j}F_{j}-\frac{1}{2}\lambda{n}_{j}^{2}\nonumber\\
   &&+\frac{1}{2}\sigma\overline{\psi}_{j}\psi_{j}+\overline{\xi}
  \psi_{j}n_{j}+\frac{N}{2g^{2}}\lambda
\end{eqnarray}
We can see that $\lambda, \xi,$ and $\sigma$ are the respective
Lagrange multiplier for the constraints:
\begin{eqnarray}
  \label{what}
  n_{j}n_{j}&=&\frac{N}{g^{2}}\nonumber\\
  n_{j}\psi_{j}&=&0\nonumber\\
  n_{j}F_{j}&=&\frac{1}{2}\overline{\psi}_{j}\psi_{j}
\end{eqnarray}
The second and the third constraints of (\ref{what})
 are supersymmetric 
transformations of the first.
We must not include kinetic terms for the field $\sigma$ and $\xi$
so as to keep these constraints manifest.
We can examine these constraints in a way that we did
in the previous section.

(1) Scalar part\\
\begin{equation}
  \label{1}
  <n_{j}n_{j}>|_{m_{n}=<\lambda>+<\sigma^{2}>}=\frac{N}{g^{2}}
\end{equation}

In two dimensions, this relation induces nonzero value to the 
mass term of the field $\vec{n}$.
\begin{eqnarray}
  m_{n}&=&<\lambda>+<\sigma>^{2}\nonumber\\
  &=&\Lambda^{2}exp\left(-\frac{4\pi}{g^{2}}\right)
\end{eqnarray}
When D=3, $m_{n}$ is nonzero in the region $g>g_{cr}$.
The critical coupling is defined by
\begin{equation}
   \frac{1}{g^{2}_{cr}}
  \equiv\int\frac{d^{3}p}{(2\pi)^{3}}\frac{1}{p^{2}}.
\end{equation}
O(N) symmetry is expected to be spontaneously broken 
by non-zero value of $v$ in the region $g<g_{cr}$. 
And when $g=g'_{cr}$, $m_{n}$ would vanish.

(2) Fermionic part\\
\begin{equation}
  \label{2}
  <n_{j}F_{j}>=\frac{1}{2}<\overline{\psi}_{j}{\psi}_{j}>
\end{equation}

This relation includes auxiliary field $F_{j}$, to be
eliminated by equation of motion.
After substituting $F_{j}$ by $\sigma{n}_{j}$, we obtain
at one-loop level: 
\begin{eqnarray}
  <n_{j}F_{j}>&=&<\sigma n_{j}n_{j}>\nonumber\\
  &=&<\sigma><n_{j}n_{j}>\nonumber\\
  &=&\frac{1}{2}<\overline{\psi}_{j}\psi_{j}>
\end{eqnarray}
If we impose the O(N) symmetric constraint 
$<n^{2}>=\frac{N}{g^{2}}$, we have
\begin{eqnarray}
  \frac{N}{g^{2}}<\sigma>&=&\frac{1}{2}
  <\overline{\psi_{j}}\psi_{j}>|_{
    m_{\psi}=<\sigma>}\nonumber\\
  \frac{N}{g^{2}}&=&N\int\frac{dp^{D}}{(2\pi)^{D}}
  \frac{1}{p^{2}+<\sigma>^{2}}.
\end{eqnarray}
For D=2, the solution is
\begin{equation}
  \label{2kai}
  <\sigma>^{2}=\Lambda^{2}exp\left(-\frac{4\pi}{g^{2}}\right).
\end{equation}
Substituting $<\sigma>$ in the first constraint (\ref{1}) with 
(\ref{2kai}), we can find that $<\lambda>$ 
must vanish.(in this point our result is different from \cite{alv}) 
This means that the field $\psi$ gains the same mass as $n$,
and simultaneously supersymmetric order parameter $<\lambda>$ 
vanishes.
We can say that the supersymmetry is not broken in two dimensions
as is predicted by Witten\cite{index}.
Moreover, we can examine the assumption of vanishing $v$  
 as follows.
We can show that the following relation can exist for
the effective potential\cite{maha}.
\begin{eqnarray}
  \label{sv}
  \frac{\partial V}{\partial v}&=&Nv(\lambda_{0}+<\sigma>^{2})
  \nonumber\\
  &=&0
\end{eqnarray}
This means that $v$ must vanish if chiral condensation
occurs. 

For D=3, we have a critical coupling constant.
As far as $g\geq{g}_{cr}$, we have nothing to worry about.
In the strong coupling region, both supersymmetry and 
O(N) symmetry are preserved in a fashion like two dimensions. 
However, in the weak coupling region, something goes wrong.
There is no non-trivial solution for constraint (\ref{2}) and 
there is no fermionic condensation (This means that the only
possible solution is $<\sigma>=0$).
Thus we can see from eq.(\ref{sv}) that $v$ can be non-zero in this
weak coupling region.
This is supported by the constraint (\ref{1}) because this 
does not have any solution in the weak coupling region
unless we allow $v$ not to vanish. 
Eq.(\ref{v}) suggests:

\begin{equation}
  v^{2}=1-g^{'2}_{cr}\int\frac{d^{3}p}{(2\pi)^{2}}\frac{1}{p^{2}}
\end{equation}

Naive consideration also supports this analysis.
In general, we can expect that quantum effects in 
correlation functions like 
$<n_{j}n_{j}>$ or $<\overline{\psi}\psi>$
would vanish in the weak coupling limit. 
But we have an O(N) symmetric constraint.
It is natural to think that the field $n$ itself gains 
 expectation value to complement quantum effects.
This simply means  that  classical effects become dominant
in the weak coupling region, therefore the O(N) symmetric
constraint is satisfied classically.
(i.e. in the weak coupling limit $g\rightarrow0$ we obtain 
$v=\pm 1$.
This is a classical solution of the constraint.)
As a result, in the weak coupling region, O(N) symmetry is
spontaneously broken by non-zero value of $v$.

We should also note that there is a possible solution of non-zero
$\lambda_{0}$.
(We neglect eq.(\ref{sv}) for a while because $\lambda_{0}$
may become a function of $v$.)
It induces a supersymmetry
breaking term to the Lagrangian:
\begin{equation}
  L_{break}=\lambda_{0}\left((\vec{n})^{2}-\frac{N}{g^{2}}
  \right)
\end{equation}
On the constrained phase 
space$\left((\vec{n})^{2}-\frac{N}{g^{2}}=0\right)$, 
vacuum energy also 
seems to vanish for non-zero $\lambda_{0}$ as long as 
the constraint (\ref{1}) is satisfied.
Does it mean there is a flat direction along $\lambda$?
Of course this statement is unnatural.
After including effective kinetic term($\sim\lambda\lambda$),
 we can find positive vacuum energy for the 
supersymmetry breaking phase.
Therefore in the supersymmetric model, $v$ is not a free parameter
but fixed by the requirement of vanishing $\lambda_{0}$.
This means $g'_{cr}$ should be adjusted to $g$, and
$v$ is fixed:
\begin{equation}
  v^{2}=1-g^{2}\int\frac{d^{3}p}{(2\pi)^{2}}\frac{1}{p^{2}}
\end{equation}

So we can conclude: 

(1) In two dimensions, both supersymmetry and O(N) symmetry 
are not broken. 
This means that  $\lambda$  and $v$ remain zero for all
value of $g$.

(2) In three dimensions, both supersymmetry and O(N) symmetry 
are not broken (i.e. $\lambda$  and $v$ remain zero) in the strong 
coupling region.
O(N) symmetry can be broken in the weak coupling region, but
supersymmetry is kept unbroken in both phases.

\section{Negative Energy}
\hspace*{\parindent}
In this section, we will reconsider whether negative energy states
in supersymmetric theories\cite{kazama,nege} can exist or not.
One may wonder why such a state appears, but it is really a 
confusing matter.
Because we have not enough space, we refer \cite{kazama} in which
detailed analysis on this topic can be found.
In ref.\cite{kazama}, two dimensional supersymmetric non-linear sigma 
model and supersymmetric Yang-Mills model are analyzed.

For us, the main problem is the value of $\lambda$.
Naively  calculated 1-loop effective 
potential 
shows that it has negative energy state at $\lambda\ne0$.
For example in D=3\cite{maha},
\begin{eqnarray}
  V&=&\frac{N}{2}\left[ \lambda_{0}\left(v^{2}-\frac{1}{g^{2}}\right)
  +v^{2}<\sigma>^{2}+\int\frac{d^{3}k}{(2\pi)^{3}}ln(k^{2}+\lambda_{0}
  +<\sigma>^{2})\right.\nonumber\\
  &&\left.-\int\frac{d^{3}k}{(2\pi)^{3}}trln(-\not{\! k}+<\sigma>)\right]
\end{eqnarray}
We can think that this problem comes 
from the instability of the tree level potential
$V=\lambda(n^{2}-N/g^{2})$ along the direction of $\lambda$.
In general  we have to set $\lambda=0$ by {\it hand},
but it should be {\it determined} by considering some effects. 

First, we are going to examine two dimensional 
non-linear sigma model from a different point of view.

We can calculate an effective potential for $\lambda$ in
two dimensional
O(N) supersymmetric non-linear sigma model using trace anomaly 
equation\cite{kazama}.
As a result, we can obtain:
\begin{equation}
  \label{kaz}
  V(\lambda)= \frac{\lambda}{8\pi}N\left[
  ln\left(\pm\frac{\Lambda^{2}}{\lambda}\right)
  +const.\right]
\end{equation}
This potential has unnatural characteristics like
negative energy or unstable vacuum.

This term can appear in the effective action at
1-loop level (we should note the meaning of $\lambda$ is 
somewhat different from eq.(\ref{kaz})).
Not yielding to a trace anomaly equation, after integrating
over $\vec{n}$ we can obtain:
\begin{eqnarray}
  \label{eff1}
  Z&=&\int D\lambda D\psi D\sigma exp\left[
  \int d^{2}x\left\{\frac{i}{2}
  \overline{\psi}_{j}\not{\! \partial}\psi_{j}
   +\frac{1}{2}\sigma\overline{\psi}_{j}\psi_{j}
   -\frac{1}{2}\sigma^{2}
   +\frac{N}{2g^{2}}\lambda-\frac{N}{2}trln(-\partial^{2}
   +\lambda+\sigma^{2})\right\}\right]\nonumber\\
 &&
\end{eqnarray}
Integration can be done for the last term 
 and we can obtain the same result as (\ref{kaz}) except for 
$\sigma^{2}$ which appeared in the mass term.

But there are some problems.
First, the effective action we derived does not 
include fermionic loop corrections that leads to the
chiral condensation.
Including the fermionic loop corrections, we can reach
at the result we have obtained  in the previous section.
The vacuum state is supersymmetric and there is no 
negative vacuum energy.
To simplify the argument, it is very useful to separate 
every constraint and discuss each property as we have done 
in section 4.

Second problem is the treatment of the effective action.
Usually we think that after integrating out $n$  fields
the integration over $\lambda$  cannot
be done exactly so we always
 consider a stationary phase approximation.
To actually determine the stationary point, we vary with
respect to the constant value of $\lambda$.
The resulting equation is the gap equation:
\begin{equation}
  \label{ab}
  \frac{N}{g^{2}}=N\int \frac{d^{2}p}{(2\pi)^{2}}
  \frac{1}{p^{2}+\lambda_{0}+<\sigma>^{2}}
\end{equation}
But there is a problem.
$\lambda$ is a Lagrange multiplier so its tree level potential 
is not stable for $\lambda$.
Naively calculating the 1-loop potential, we will find
(fictitious) negative energy state.
In general supersymmetric non-gauge theories, $F_{j}F_{j}$ 
type term 
in the kinetic term($\Phi_{j} D^{2} \Phi_{j}$) is 
responsible for the positivity of the vacuum energy.
With this term, scalar potential is always written as
$V=\sum |W_{i}|^{2}$.  
In our model, $\lambda\lambda$ term will appear in
the effective kinetic term and is responsible for the
positivity of the vacuum energy.
Of course, there is a possibility that the kinetic term 
would be a non-trivial(special) function of $\Sigma$.
Then, the positivity of the vacuum energy is not manifest and
the argument of negative energy would be trustworthy.
(But in our approximation such a term does not appear.)
Moreover, as we have shown in the previous section,
the stationary point $\lambda_{0}$ 
is exactly determined by fermionic constraint in two dimensions
and resulting effective potential $V^{eff}(n_{j})$
 vanishes  in the stationary phase
approximation.

Can we apply the same argument to the three dimensional model?
Naively calculating the 1-loop effective potential,
a negative energy state appears in the wrong vacuum
$\lambda\ne0$ even if we consider the fermionic condensation.
In this case, we must also consider the effective kinetic term that 
yields effective $\lambda\lambda$ term.
Including this, we can expect that the scalar potential 
is always positive.

\section{Conclusion}
\hspace*{\parindent}
Some authors claimed that in supersymmetric models,
there can be a supersymmetry
breaking accompanied by
 negative energy and negative norm states that 
lead to other instabilities.
It is true that we cannot ignore such a possibility in general
but we can make sure of the absence of such a vacuum
at least in $O(N)$ sigma model in two and three dimensions.

Merely adding the Lagrange multiplier fields and taking it
as a scalar potential, we would be led to  unnatural arguments.
If relating the Lagrange multiplier to the potential is necessary, 
we should have considered about the effective kinetic terms.
Of course, we must be careful not to forget to include both 
fermionic and bosonic loops\cite{kazama,mat}.
The same can be said for the analysis of  supersymmetric
 Yang-Mills or supersymmetric QCD theories.
Decomposed in component fields, these theories look like
ordinary QCD with Majorana fermions or that with Higgs fields.
So we tend to forget their origin and analyze these theories in 
usual way of QCD.

We have analyzed the phase structures of O(N) supersymmetric 
sigma model in two and three dimensions by using the tadpole method. 

We have shown that after including fermionic constraint and
a effective kinetic term,
 $\lambda$ is {\it determined} as $\lambda=0$ and 
the supersymmetry breaking vacuum has positive energy.
There is no fear of negative energy states 
at least in $O(N)$ sigma model discussed above.

\section*{Acknowledgment}
We thank K.Fujikawa, Y.Kazama,
 S.Iso and K.Hori  for many helpful discussions.

\end{document}